\documentclass[a4paper]{cas-dc}
\usepackage[numbers,sort&compress]{natbib}
\usepackage{braket}
\usepackage{multicol}

\definecolor{darkpastelgreen}{rgb}{0.01, 0.75, 0.24}
\definecolor{electricindigo}{rgb}{0.44, 0.0, 1.0}
\definecolor{palatinateblue}{rgb}{0.15, 0.23, 0.89}
\definecolor{carminered}{rgb}{1.0, 0.0, 0.22}
\hypersetup{linkcolor=palatinateblue, citecolor=electricindigo, urlcolor=carminered}

\begin{document}
\shorttitle{Dirac/Weyl-node-induced oscillating Casimir effect}
\shortauthors{K. Nakayama and K. Suzuki}  
\title[mode = title]{Dirac/Weyl-node-induced oscillating Casimir effect}  

\author[1,2]{Katsumasa Nakayama}[type=editor, orcid=0000-0003-0270-8523]
\nonumnote{$\dag$ E-mail: katsumasa.nakayama@riken.jp (corresponding author)}

\author[3]{Kei Suzuki}[orcid=0000-0002-8746-4064]
\nonumnote{$\ddag$ E-mail: k.suzuki.2010@th.phys.titech.ac.jp (corresponding author)}

\address[1]{RIKEN Center for Computational Science, Kobe 650-0047, Japan}
\address[2]{NIC, DESY Zeuthen, Platanenallee 6, 15738 Zeuthen, Germany}
\address[3]{Advanced Science Research Center, Japan Atomic Energy Agency (JAEA), Tokai, 319-1195, Japan}

\begin{abstract}
The Casimir effect is a quantum phenomenon induced by the zero-point energy of relativistic fields confined in a finite-size system.
This effect for photon fields has been studied for a long time, while the realization of counterparts for fermion fields in Dirac/Weyl semimetals is an open question.
We theoretically demonstrate the typical properties of the Casimir effect for relativistic electron fields in Dirac/Weyl semimetals and show the results from an effective Hamiltonian for realistic materials such as Cd$_3$As$_2$ and Na$_3$Bi.
We find an oscillation of the Casimir energy as a function of the thickness of the thin film, which stems from the existence of Dirac/Weyl nodes in momentum space.
Experimentally, such an effect can be observed in thin films of semimetals, where the thickness dependence of thermodynamic quantities is affected by the Casimir energy.
\end{abstract}


\maketitle

\section{Introduction}
The well-known Casimir effect~\cite{Casimir:1948dh,Lamoreaux:1996wh,Bressi:2002fr} is a quantum phenomenon for photon fields between two parallel plates with nanoscale distance, which results in mechanical/thermodynamic effects, such as the Casimir force and Casimir pressure (see Refs.~\cite{Plunien:1986ca,Mostepanenko:1988bs,Bordag:2001qi,Milton:2001yy,Klimchitskaya:2009cw} for reviews).
Thus, the Casimir effect for photons may play an important role in the field of nanophotonics~\cite{Gong:2020ttb}.

\begin{figure}[t!]
    \centering
    \begin{minipage}[t]{1.0\columnwidth}
    \includegraphics[clip,width=1.0\columnwidth]{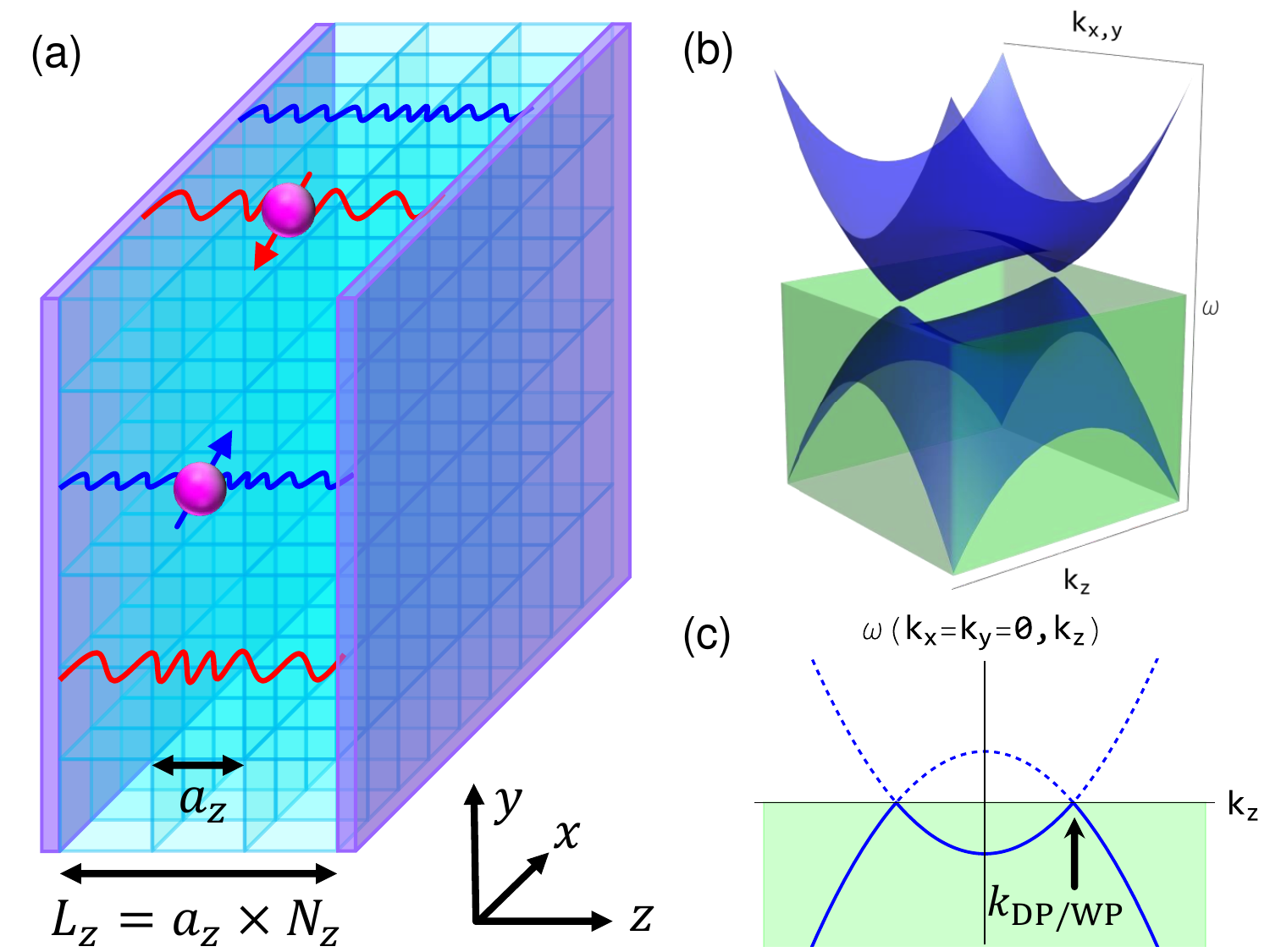}
    \end{minipage}
    \caption{(a) Schematic picture of Casimir effect in thin films of three-dimensional materials.
(b) An example of energy-momentum dispersion relation $\omega({\bf k})$ for electrons in Dirac/Weyl semimetals.
(c) Dispersion relation $\omega(k_z)$ for the $z$ component of momentum at $k_x=k_y=0$.
The Dirac or Weyl points are located at ${\bf k} = (0,0,\pm k_\mathrm{DP/WP})$.
}
    \label{schematic}
\end{figure}

In general, the Casimir effect is not limited to photonic systems, and the understanding of fermionic counterparts is an important and open issue.
For example, the Casimir effect for free electron fields in vacuum is suppressed by the electron mass, so that its observation is difficult.
On the other hand, massless fermions are realized in solid-state systems such as Dirac semimetals (DSMs)~\cite{Murakami:2007,Young:2012,Wang:2012,Wang:2013} and Weyl semimetals (WSMs)~\cite{Wan:2010fyf,Yang:2011,Burkov:2011ene} (see Ref.~\cite{Armitage:2017cjs} for a review).
Small-size systems such as thin films of three dimensional (3D) material [see Fig.~\ref{schematic}(a)], narrow nanoribbons, and short nanowires may induce Casimir effects for fermion fields.
The Casimir energy is a part of the free energy (or thermodynamic potential) of the system and should contribute to the internal pressure and other quantities. 
In such materials, if electrons carry electric or spin degrees of freedom, the understanding of the electronic Casimir effect could be essential for electronics and spintronics using thin films.
In Table~\ref{tab:comp}, we emphasize the comparison between the conventional works and our study.\footnote{The purpose of our paper should be distinguished from {\it photonic} Casimir effects between topological insulators~\cite{Grushin:2010qoi,Grushin:2011,Chen:2011,Babamahdi:2021cdk}, between Chern insulators~\cite{Tse:2012pb,Rodriguez-Lopez:2013pza,Fialkovsky:2018fpo}, between Dirac/Weyl semimetals~\cite{Wilson:2015,Rodriguez-Lopez:2019oex,Chen:2020ova,Farias:2020qqp,Bordag:2021dxm}, and inside chiral materials~\cite{Jiang:2018ivv,Fukushima:2019sjn,Canfora:2022xcx} (see Refs.~\cite{Woods:2015pla,Lu:2021jvu} for reviews).}

\begin{table}[t!]
  \centering
  \caption{Comparison of Casimir effects.}
\begin{tabular}{l l l } 
  \hline \hline
          & Conventional works & This work \\ 
\hline
Field           & photons & Dirac/Weyl electrons \\
System       & QED vacuum & bulk of thin film \\
Boundary    & two plates, etc. & edge of thin film \\
Space         & continuous & lattice \\
Applications & photonics & electronics/spintronics \\
\hline \hline   
  \end{tabular}
  \label{tab:comp}
\end{table}

In this Letter, we theoretically investigate the Casimir effect originated from Dirac/Weyl fermion fields inside thin films of 3D DSMs/WSMs.
In particular, we point out an importance of the Dirac points (DPs) or Weyl points (WPs) that exists at a nonzero ${\bf k} =(k_x,k_y,k_z)$ in momentum space, as shown in Figs.~\ref{schematic}(b) and (c).
We find that these points (or nodes) induce a novel type of Casimir effect: An oscillation of Casimir energy as a function of the film thickness, which we call the {\it Dirac/Weyl-node-induced oscillating Casimir effect}.
Note that this phenomenon does not occur for conventional fields in elementary particle physics, such as photons, electrons, quarks, and neutrinos, because the DP or WP of these fields exists at the origin (namely, the $\Gamma$ point) in momentum space.
In addition, as illustrative examples of this effect, we discuss systems under an external magnetic field and realistic DSMs such as Cd$_3$As$_2$ and Na$_3$Bi.

%%%%%%%%%%%%%%%%%%%%%%%%%%%%%%%%
\section{Casimir effect on the lattice}
The Casimir effect for fields on the lattice can be defined by using a lattice regularization \cite{Actor:1999nb,Pawellek:2013sda,Ishikawa:2020ezm,Ishikawa:2020icy,Nakayama:2022ild,Nakata:2022pen,Mandlecha:2022cll}.\footnote{Numerical simulations on the lattice are another powerful approaches.
See the U(1) gauge field~\cite{Pavlovsky:2009kg,Pavlovsky:2010zza,Pavlovsky:2011qt}, the compact U(1) gauge field~\cite{Pavlovsky:2009mt,Chernodub:2016owp,Chernodub:2017mhi,Chernodub:2017gwe,Chernodub:2022izt}, the SU(2) gauge field~\cite{Chernodub:2018pmt,Chernodub:2018aix}, and the SU(3) gauge field~\cite{Kitazawa:2019otp}.}
In this work, we focus on thin films confined in the $z$ direction,\footnote{In this setup, there are no surface Fermi arcs on the $k_x$-$k_y$ plane, so that the thermodynamics of the system is dominated by the bulk fermions.
For thin films confined in the $x$ or $y$ direction, one can also calculate the Casimir energy for the bulk in a similar manner, but the contribution of the surface states should be taken into account if we consider the open boundary.} where the film thickness is $L \equiv a_zN_z$ with the lattice constant $a \equiv a_z$ and the number of lattice cells $ N_z$ [see Fig.~\ref{schematic}(a)].
The Casimir energy $E_\mathrm{Cas}$ for $N_z$ per unit area of the surface is defined as
\begin{subequations}
\begin{align}
E_\mathrm{Cas} &\equiv E_0^\mathrm{sum}(N_z) - E_0^\mathrm{int}(N_z) \label{eq:def_cas} \\
E_0^\mathrm{sum}(N_z) &= \sum_j \int_\mathrm{BZ} \frac{d^2(a_ik_\perp)}{(2\pi)^2} \left[ - \frac{1}{2}  \sum_{n} |\omega_{k_\perp,n,j}|  \right], \\
E_0^\mathrm{int}(N_z) & =  \sum_j \int_\mathrm{BZ} \frac{d^2(a_ik_\perp)}{(2\pi)^2} \left[ -\frac{N_z}{2} \int_\mathrm{BZ} \frac{d(a_zk_z)}{2\pi} |\omega_{{\bf k},j}| \right],
\end{align}
\end{subequations}
$E_0^\mathrm{sum}$ and $E_0^\mathrm{int}$ are the zero-point energies consisting of discrete energies $\omega_{k_\perp,n,j}$ and continuous energies $\omega_{{\bf k},j}$, respectively.
In this work, we apply a phenomenological boundary condition,\footnote{
In condensed matter physics, such an approximate condition is often used (e.g., see Refs.~\cite{Xiao:2015,Pan:2015}).
In this condition, fermion fields are regarded as nonrelativistic fields bounded by a infinite potential well in the $z$ direction.
For continuous massive Dirac fermions under the MIT bag boundary conditions~\cite{Johnson:1975zp}, the heavy-mass limit also leads to the same eigenvalue structure.} where $k_z$ in $E_0^\mathrm{sum}$ is discretized as $a_zk_z \to \frac{n\pi}{N_z}$ ($n=1,\cdots, 2N_z$), and $\sum_{n} \to \frac{1}{2} \sum_n$.
The momentum integral is within the first Brillouin zone (BZ).
The transverse momentum is defined as $k_\perp^2 \equiv k_x^2+k_y^2$, and $d^2(a_ik_\perp) \equiv d(a_xk_x)d(a_yk_y)$.
$j$ labels all the eigenvalues, and the absolute value is needed for hole degrees of freedom.
The overall factor of $-1/2$ comes from the zero-point energy of fermion fields.

%%%%%%%%%%%%%%%%%%%%%%%%%%%%%%%%
\section{Casimir effect in WSMs} \label{Sec:3}
\begin{figure}[tb!]
    \centering
    \begin{minipage}[t]{0.33\columnwidth}
    \includegraphics[clip,width=1.0\columnwidth]{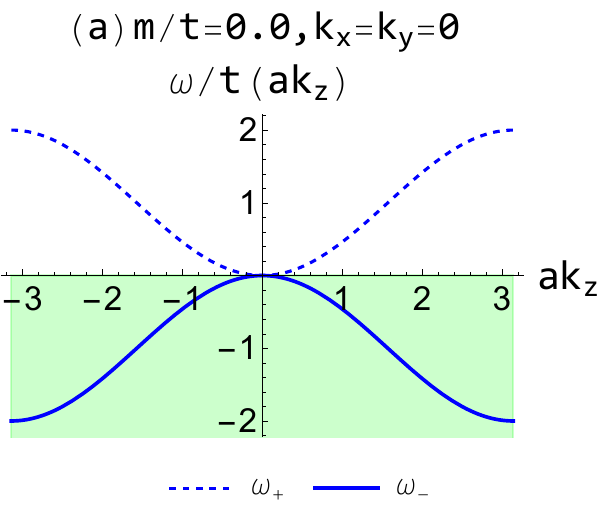}
    \end{minipage}%
    \begin{minipage}[t]{0.33\columnwidth}
    \includegraphics[clip,width=1.0\columnwidth]{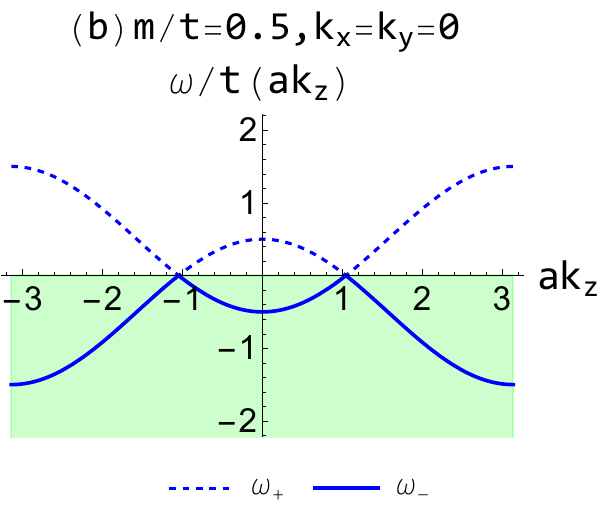}
    \end{minipage}%
    \begin{minipage}[t]{0.33\columnwidth}
    \includegraphics[clip,width=1.0\columnwidth]{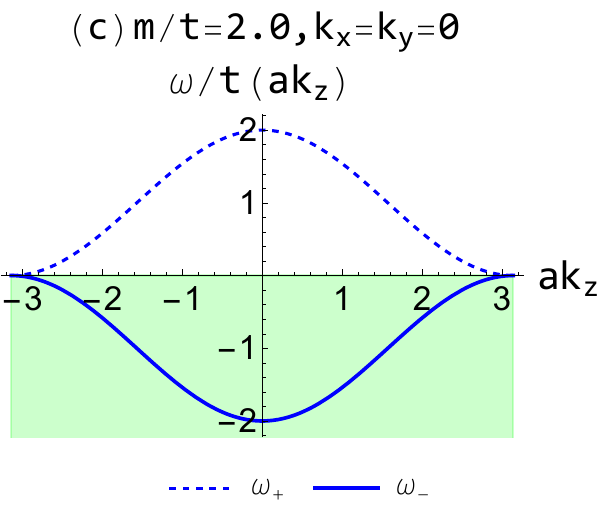}
    \end{minipage}
    \begin{minipage}[t]{1.0\columnwidth}
    \includegraphics[clip,width=1.0\columnwidth]{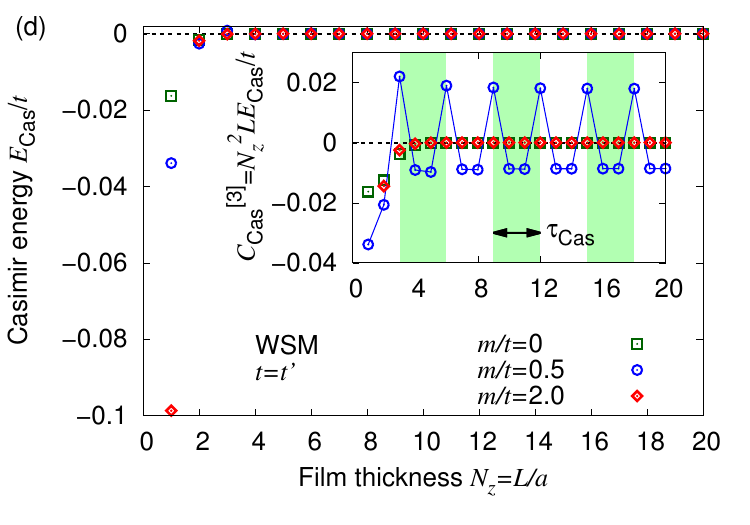}
    \end{minipage}
    \caption{Dispersion relations, Casimir energy, and Casimir coefficient for Weyl fermions defined in the toy model~(\ref{eq:WSMtoy}).
(a)-(c) Dispersion relations $\omega(k_z)/t$ in the $z$ direction, where the model parameter is fixed as $m/t=0.0$, $0.5$, or $2.0$.
(d) Casimir energy $E_\mathrm{Cas}/t$ and Casimir coefficient $C_\mathrm{Cas}^{[3]}/t$ for (001)-oriented thin films.
Inset: the period of $\tau_\mathrm{Cas} = 3$ for $m/t=0.5$ is shown as the colored or uncolored regions.}
    \label{toy}
\end{figure}

First, we demonstrate the typical behaviors of the Casimir effect for Weyl fermions in thin films of WSMs.
By diagonalizing an effective Hamiltonian~\cite{Yang:2011} for time-reversal broken WSMs (also see Appendix \ref{App:1}), we obtain the typical dispersion relations for Weyl electrons:
\begin{align}
\omega_\pm^\mathrm{WSM} = \pm \sqrt{ t^2 \sum_i^{x,y} \sin^2 a_ik_i + \left[m - t^\prime \sum_i^{x,y,z} (1-\cos a_ik_i) \right]^2}. \label{eq:WSMtoy}
\end{align}
Below, we fix $t^\prime= t$ and focus on $m/t = 0$, $0.5$, and 2.0.
The dispersion relations are shown in Figs.~\ref{toy}(a)-(c).

In Fig.~\ref{toy}(d), we show the numerical results of the dimensionless Casimir energies $E_\mathrm{Cas}/t$.
For all $m$, we find nonzero Casimir energies.
As the film thickness $L=aN_z$ increases, the Casimir energy is damped.
To see the detailed $N_z$ dependence of the damping, we plot Casimir coefficients~\cite{Ishikawa:2020icy} defined as $C_\mathrm{Cas}^{[3]} \equiv N_z^3E_\mathrm{Cas}/t$, where $[3]$ means the exponent of the prefactor $N_z^3$.
For example, the typical behavior for the Dirac/Weyl fermions with a linear dispersion $\omega_\pm^\mathrm{D/W} \propto \pm |{\bf k}|$ in the three-dimensional continuous space is $E_\mathrm{Cas} \propto 1/L^3$, and its coefficient $C_\mathrm{Cas}^{[3]} \equiv L^3 E_\mathrm{Cas}$ becomes constant.
As shown in the inset of Fig.~\ref{toy}(d), the Casimir coefficient at $m/t=0$ and $2.0$ converges to zero.
This behavior means that the damping of the Casimir energy is faster than $1/N_z^3$.
This is understood by the dispersion relations shown in Figs.~\ref{toy}(a) and (c):
The band-touching region near the Fermi level is quadratic-like, which leads to faster damping.\footnote{
This damping is similar to the behavior for a massive field in the continuum theory (e.g., \cite{Hays:1979bc,Mamaev:1980jn,Ambjorn:1981xw}).
For a consideration for quadratc dispersions on the lattice, see Ref.~\cite{Nakayama:2022ild}.}

On the other hand, for $m/t=0.5$ as shown in Fig.~\ref{toy}(b), there exist the pair of WPs and linear dispersions, and hence the Casimir effect is qualitatively different:
We find a clear oscillation of the Casimir energy when $N_z$ is large enough.
Since this behavior is caused by the existence of the Weyl points (or nodes) at the Fermi energy,\footnote{A similar oscillation of the Casimir energy was predicted for the naive lattice fermion, which is the same form as Eq.~(\ref{eq:WSMtoy}) replacing as $\sum_i^{x,y} \to \sum_i^{x,y,z}$ at $t\neq 0$ and $t^\prime=0$, with the periodic or antiperiodic boundary condition~\cite{Ishikawa:2020ezm,Ishikawa:2020icy}.
Furthermore, for ``physical" boundary conditions such as the MIT bag boundary~\cite{Johnson:1975zp}, this oscillation is absent~\cite{Mandlecha:2022cll}.
Unlike the naive fermion, the oscillation predicted in the present work survives not only for the periodic and antiperiodic boundary conditions but also for physical boundary conditions.} we call it the node-induced oscillating Casimir effect.
This oscillation is induced by the discretization of the momentum $k_z$ on the dispersion relations within the BZ, so that the period $\tau_\mathrm{Cas}$ of the oscillation can be connected to the positions ${\bf k} = (0,0, \pm k_\mathrm{WP/DP})$ of the WPs (or DPs):\footnote{The formula~(\ref{eq:period}) depends on the boundary condition.
As an instructive example, for the periodic boundary condition, $ak_z \to \frac{2\pi n}{N_z}$ ($n=0,1,\cdots,N_z-1$), we obtain $\tau_\mathrm{Cas} = \frac{2\pi}{a_z k_\mathrm{WP/DP}}$ (see Supplementary Material S2)}
\begin{align}
\tau_\mathrm{Cas} = \frac{\pi}{a_z k_\mathrm{WP/DP}} . \label{eq:period}
\end{align}
For the parameters used here, the WPs are located at $a_z k_\mathrm{WP} = \pi/3$, so that the corresponding period is estimated as $\tau_\mathrm{Cas} = 3$, which is consistent with Fig.~\ref{toy}(d).
The formula~(\ref{eq:period}) means that the period as a function of the film thickness $L = N_za_z$ is $\frac{\pi}{k_\mathrm{WP/DP}}$, which is determined only by $k_\mathrm{WP/DP}$.\footnote{In this section, we {\it numerically} have found the oscillatory behavior and its period, while they can be also {\it analytically} derived for a simplified one-dimensional dispersion (see Supplementary Material S4).}

An intuitive interpretation for the node-induced oscillation is as follows.
The Casimir energy is caused by the difference between the integral of $E_0^\mathrm{int}$ over the continuous momentum and the sum of $E_0^\mathrm{sum}$ over the discretized momenta.
Here, $E_0^\mathrm{int}/N_z$ is constant while the contribution from $E_0^\mathrm{sum}$ around the WPs strongly depends on $N_z$.
If $a k_\mathrm{WP} = \pi/3$, for $N_z = 1,2$, the possible momenta do not match the WPs. 
On the other hand, for $N_z = 3$, the momenta are $ak_z = 0, \pm \frac{\pi}{3}, \pm \frac{2\pi}{3}, \pi$, and particularly $\pm \frac{\pi}{3}$ match the WPs.
Since, in general, the linear-dispersion region enhances the Casimir effect, the Casimir energy for $N_z = 3$ is stronger than those for $N_z = 1,2$.
Thus, when $N_z$ is a multiple of 3, the momenta at the WPs are included into the Casimir energy.
This interpretation holds for any $k_\mathrm{WP}$.

%%%%%%%%%%%%%%%%%%%%%%%%%%%%%%%%
\section{Casimir effect in WSM/DSM with a magnetic field} \label{Sec:4}
Next, we study the Casimir effect under an external magnetic field.
In this work, we apply a magnetic field parallel to the $z$ axis, where the continuous momenta $k_x$ and $k_y$ are replaced by discretized levels as Landau levels (LLs) (also see Appendix \ref{App:2}).
The dispersion relation of the zeroth LL (0LL) for the spinless Weyl fermion is given by~\cite{Nguyen:2021} 
\begin{align}
\omega^\mathrm{WSM-0LL} = -m + t^\prime (1-\cos ak_z) + \pi t^\prime \phi \label{eq:LLL},
\end{align}
where $m$ and $t^\prime$ are the same model parameters as those in Eq.~(\ref{eq:WSMtoy}).
$\phi \equiv eBa_xa_y/h$ is the magnetic flux with a magnetic-field strength $B$, the electric charge $e$, and the Planck constant $h$.
When the magnetic field is strong enough, the Casimir energy is expected to be dominated by the contribution from the zeroth LL, and we neglect higher LLs.
The dispersion relation (\ref{eq:LLL}) is written as the constant term $ -m +\pi t^\prime \phi$ and the $k_z$-dependent term, so that the tunable dimensionless parameter within this model is $ -m/t^\prime + \pi  \phi$.
Experimentally, since $m$ and $t^\prime$ are intrinsic values, one can tune the band structure by changing the strength of the magnetic field.

\begin{figure}[tb!]
    \centering
    \begin{minipage}[t]{1.0\columnwidth}
    \includegraphics[clip,width=1.0\columnwidth]{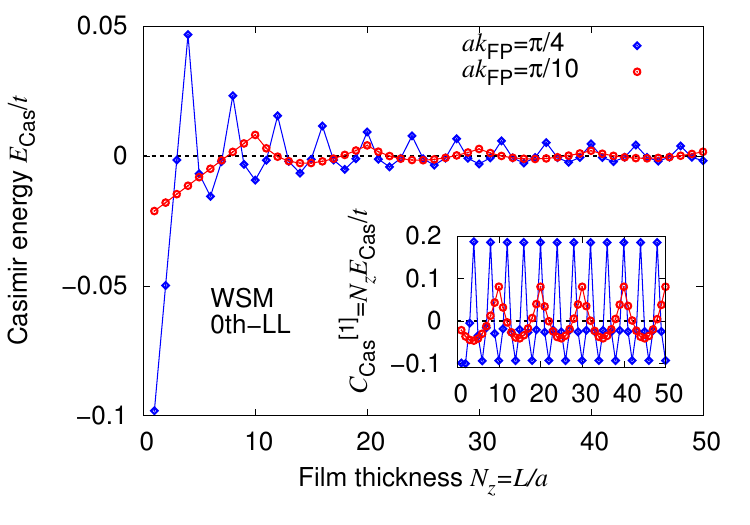}
    \end{minipage}
    \caption{Casimir energy and Casimir coefficient for the zeroth Landau level in a toy model (\ref{eq:LLL}) of Weyl semimetals in a magnetic field.
Positions of the Fermi points are located at $ak_\mathrm{FP}=\pi/4$ or $\pi/10$.}
    \label{magWSM}
\end{figure}

In Fig.~\ref{magWSM}, we show the numerical results for the zeroth LL of the Weyl fermion (\ref{eq:LLL}).
The dispersion relation (\ref{eq:LLL}) is a downward convex function, and there are two crossing points across the Fermi level, namely Fermi points (FPs).
Here, we consider the dispersion relation with the FPs at $ak_z=\pm \pi/4$ or $\pm \pi/10$, where $ -m/t^\prime + \pi  \phi$ is tuned so as to satisfy $\omega^\mathrm{WSM-0LL} =0$.
In Fig.~\ref{magWSM}, we again find the oscillation of the Casimir energies with periods, $\tau_\mathrm{Cas} = 4$ and $\tau_\mathrm{Cas} = 10$, which is consistent with the formula~(\ref{eq:period}).

\begin{figure}[tb!]
    \centering
    \begin{minipage}[t]{1.0\columnwidth}
    \includegraphics[clip,width=1.0\columnwidth]{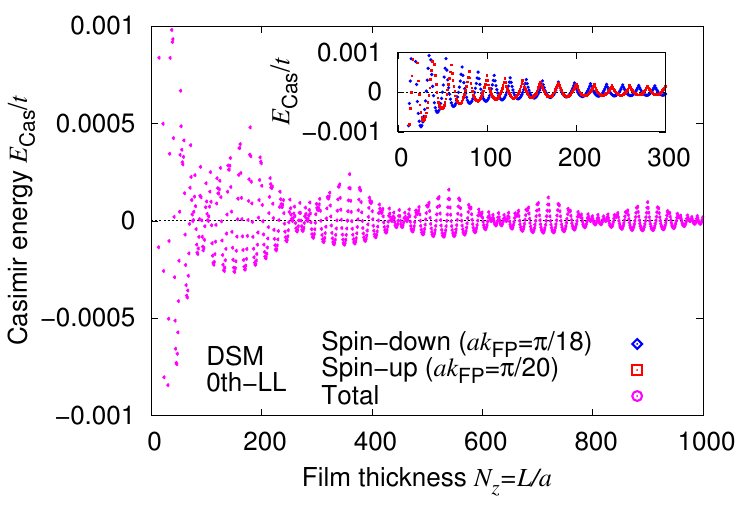}
    \end{minipage}
    \caption{Total Casimir energy for the zeroth LLs in a toy model (\ref{eq:spinup})-(\ref{eq:spindown}) of Dirac semimetal in a magnetic field.
Inset: Casimir energies for the zeroth LL of the spin-up or spin-down band.}
    \label{magDSM}
\end{figure}

For DSMs, most of the behaviors are similar to WSMs, but hereafter we demonstrate a characteristic behavior.
The spin-up and spin-down bands in DSMs are degenerate in the absence of a magnetic field.
When a magnetic field is switched on, the spin degeneracy is lifted by the Landau quantization and the Zeeman effect.
As a result, two dispersion relations in the zeroth LL are given by~\cite{Nguyen:2021}
\begin{subequations}
\begin{align}
\omega^\mathrm{DSM-0LL}_\uparrow &= m - t^\prime (1-\cos ak_z) - \pi t^\prime \phi + \lambda_z g_\uparrow \phi \label{eq:spinup}, \\
\omega^\mathrm{DSM-0LL}_\downarrow &= -m + t^\prime (1-\cos ak_z) +\pi t^\prime \phi - \lambda_z g_\downarrow \phi \label{eq:spindown},
\end{align}
\end{subequations}
where $\lambda_z \equiv \mu_Bh/2ea_xa_y$ with the Bohr magneton $\mu_B$, and the $g$ factor $g_{\uparrow(\downarrow)}$ characterizes the magnetic moment of the up-spin or down-spin band.
Thus, the band splitting induced by the Landau quantization and the Zeeman effect can lead to a characteristic behavior of the Casimir effect.
In Fig.~\ref{magDSM}, we show the total Casimir energy for the zeroth LLs including both the spin-up and spin-down bands, and the plots for the two bands are shown separately in the inset. 
Here, we choose parameters so that the periods for two bands are slightly different ($\tau_\mathrm{Cas} = 20$ and $18$).
Then, for the total Casimir energy, two types of oscillations are combined and induce a ``beat" of Casimir energy, where its period is $1/(\frac{1}{18} - \frac{1}{20}) = 180$.
Such a beating behavior of the Casimir effect, which is a periodic enhancement and suppression, will be useful for tuning the Casimir effect in DSMs by applying an external magnetic field.\footnote{For a quantitative analysis with the zeroth LL of Cd$_3$As$_2$, see Supplementary Material S3.}

%%%%%%%%%%%%%%%%%%%%%%%%%%%%%%%%
\section{Casimir effect in Cd$_3$As$_2$ and Na$_3$Bi}
\begin{figure}[t!]
    \centering
    \begin{minipage}[t]{0.33\columnwidth}
    \includegraphics[clip,width=1.0\columnwidth]{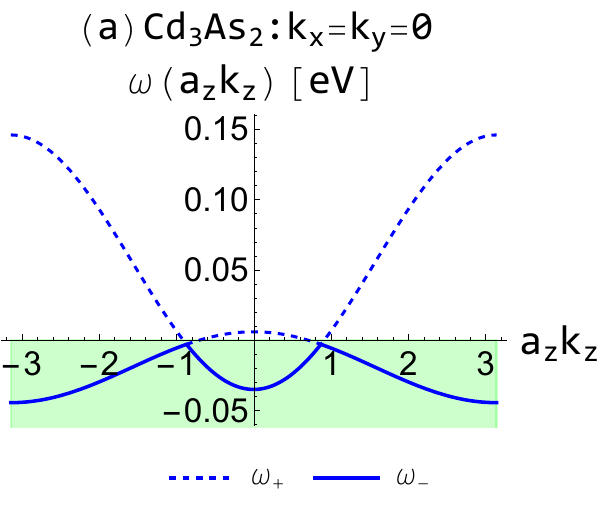}
    \end{minipage}%
    \begin{minipage}[t]{0.33\columnwidth}
    \includegraphics[clip,width=1.0\columnwidth]{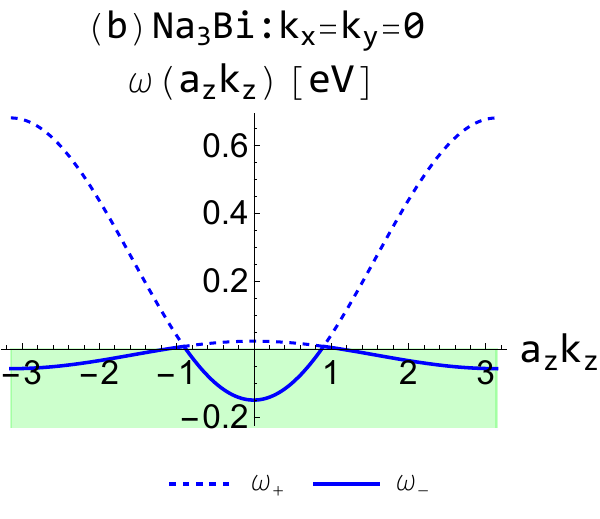}
    \end{minipage}
    \begin{minipage}[t]{1.0\columnwidth}
    \includegraphics[clip,width=1.0\columnwidth]{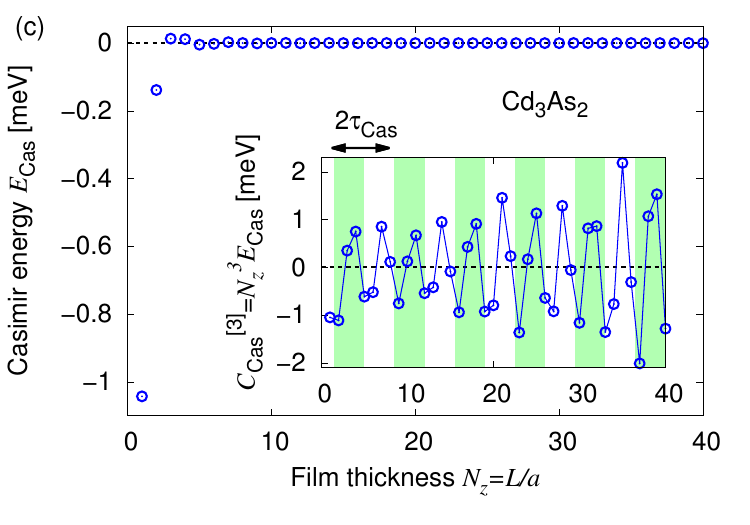}
    \end{minipage}
    \begin{minipage}[t]{1.0\columnwidth}
    \includegraphics[clip,width=1.0\columnwidth]{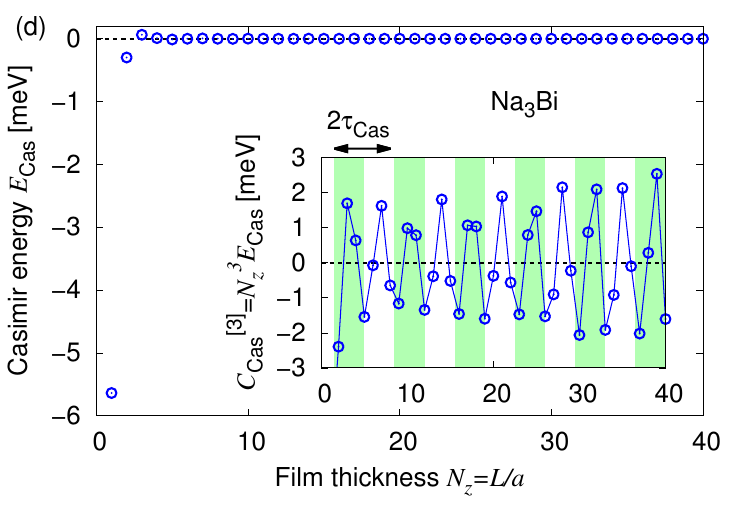}
    \end{minipage}
    \caption{Dispersion relations, Casimir energy, and Casimir coefficient for Dirac electrons in Cd$_3$As$_2$ or Na$_3$Bi thin films, described by the model~(\ref{eq:DSM}).
(a)-(b) Dispersion relations in the $z$ direction.
(c)-(d) Casimir energy and Casimir coefficient for (001)-oriented thin films.
Inset: the period of $\tau_\mathrm{Cas} \sim 3.5$ is shown as the colored or uncolored regions.}
    \label{fig:DSM}
\end{figure}

Here, we evaluate the Casimir effect for Dirac electrons realized in Cd$_3$As$_2$ and Na$_3$Bi which are regarded as 3D Dirac semimetals~\cite{Wang:2012,Wang:2013}.
For experimental evidence, see Refs.~\cite{Neupane:2014,Borisenko:2014,Liu:2014,Jeon:2014,He:2014,Yi:2014bla} for Cd$_3$As$_2$, Refs.~\cite{Liu:2015,Schumann:2018,Uchida:2017,Galletti:2018,Kealhofer:2019,Kealhofer:2020,Goyal:2020,Kealhofer:2021} for Cd$_3$As$_2$ thin films, Refs.~\cite{Liu:2014Science,Xu:2015} for Na$_3$Bi, and Refs.~\cite{Zhang:2014,Collins:2018,Bernardo:2021} for Na$_3$Bi thin films.
In particular, our setup is suitable for (001)-oriented thin films of Cd$_3$As$_2$~\cite{Kealhofer:2019,Kealhofer:2020,Goyal:2020,Kealhofer:2021}.
A low-energy effective Hamiltonian near the DPs was proposed in Refs.~\cite{Wang:2012,Wang:2013} (also see Supplementary Material S1), where the four-band dispersion relations (the spin-up and spin-down bands are degenerate) are given by
\begin{align}
\omega_\pm^\mathrm{DSM} = \epsilon_0 \pm \sqrt{M^2+A^2(k_x^2+k_y^2)}, \label{eq:DSM}
\end{align}
where $\epsilon_0 = C_0 + C_1k_z^2 + C_2(k_x^2+k_y^2)$ and $M = M_0 +M_1k_z ^2 +M_2 (k_x^2+k_y^2)$.
See Table~\ref{tab:model_para1} for the model parameters.
In lattice space, we replace the momentum as $k_i^2 \to \frac{1}{a_i^2} \sin^2 ak_i$ for the term proportional to $A$ and $k_i^2 \to \frac{1}{a_i^2} (2-2\cos ak_i)$ for the other terms.
The dispersion relations are shown in Figs.~\ref{fig:DSM}(a) and (b).
Note that these dispersion relations are reliable only near the DPs, but it is enough to discuss qualitative behaviors of the Casimir effect.

\begin{table}[b!]
  \centering
  \caption{Model parameters for Cd$_3$As$_2$~\cite{Cano:2016eie,Steigmann:1968} and Na$_3$Bi~\cite{Wang:2012}.}
\begin{tabular}{l c c c c} 
  \hline \hline
Parameters   & Cd$_3$As$_2$~\cite{Cano:2016eie,Steigmann:1968} & Na$_3$Bi~\cite{Wang:2012} \\ 
  \hline
$A$ (eV\AA)            &  $0.889$     & $2.4598 $ \\
$C_0$ (eV)               &  $ -0.0145$ & $-0.06382$ \\
$C_1$ (eV\AA$^2$)  &   $10.59$     & $8.7536$  \\
$C_2$ (eV\AA$^2$)  &   $11.5$       & $ -8.4008$ \\
$M_0$ (eV)              &   $ -0.0205$ & $ -0.08686$\\
$M_1$ (eV\AA$^2$)  &   $18.77$     & $10.6424$ \\
$M_2$ (eV\AA$^2$)  &   $13.5$      & $10.361$ \\
$a_x=a_y$ (\AA)        &  $12.67$     & $ 5.448$ \\
$a_z$ (\AA)              &  $25.48$     & $9.655$ \\
  \hline \hline   
  \end{tabular}
  \label{tab:model_para1}
\end{table}

In Figs.~\ref{fig:DSM}(c) and (d), we show the numerical results for the Casimir energy and Casimir coefficient.
From these figures, we find an oscillation of the Casimir energy.
This oscillation is determined by the complicated band structures including not only the two DPs but also the four FPs.\footnote{This is a simplified interpretation focusing on the Fermi points along the $k_z$ axis at $k_x=k_y=0$.
Precisely speaking, the band structures across the Fermi energy at $k_x \neq 0$ and/or $k_y \neq 0$ also contribute to the Casimir effect.}
Here, a DP and two FPs are close enough, so that the oscillation period is roughly determined by only the positions of DPs.
From the model parameters, we can roughly estimate $a_z k_\mathrm{DP} \sim \sqrt{-a_z^2M_0/M_1} \sim \pi/3.73$ and $\pi/3.60$ for Cd$_3$As$_2$ and Na$_3$Bi, respectively.
By substituting these values into the formula~(\ref{eq:period}), we expect $\tau_\mathrm{Cas} = \pi/a_z k_\mathrm{DP} \sim 3.73$ and $3.60$, respectively.
These estimates are consistent with the numerical results: we can find $\tau_\mathrm{Cas} \sim 3.5$.
For example, when $N_z$ is a multiple of 7, $C_\mathrm{Cas}^{[3]}$ is stronger than the values at other $N_z$.
Thus, the formula~(\ref{eq:period}) is useful even for realistic materials. 
Experimentally, one can change the film thickness $L \equiv a_zN_z$, and the oscillation periods for Cd$_3$As$_2$ and Na$_3$Bi are expected to be $L = a_z\tau_\mathrm{Cas} \sim 9.5$ nm and $3.5$ nm, respectively.

Note that, as shown in Refs.~\cite{Wang:2012,Wang:2013,Xiao:2015,Pan:2015,Chen:2017,Chen:2021}, the discretization of $k_z$ in DSM thin films is regarded as ``energy gaps" opened at the DPs, and hence topological phase transition as a function of the film thickness can occur.
For instance, a thickness-dependent periodicity of the topological invariant in Cd$_3$As$_2$ thin films was shown in Fig.~4 in Ref.~\cite{Wang:2013}.
This periodicity corresponds to that of the Casimir energy.
Thus, the Casimir effect is another phenomenon induced by the discretization of $k_z$ and is regarded as an indirect signal of topological phase transition.
Also, we emphasize that even if the Fermi level is slightly shifted or gaps are opened by external perturbations, such as strain~\cite{Arribi:2020}, the node-induced oscillation is qualitatively unchanged (see Supplementary Material S1).

%%%%%%%%%%%%%%%%%%%%%%%%%%%%%%%%
\section{Experimental observables}
Finally, we comment on experimental realization of the Casimir effect in DSMs/WSMs.
In experiments, the thickness of thin films ($L \propto N_z$, where $a_z$ is usually fixed) is controllable by microfabrication technology, and one can investigate the thickness dependence of physical quantities.
The Casimir energy is a part of the free energy which survives even at zero temperature, so that it intrinsically contributes to the free energy inside the thin film.
Quantitatively, within our model, the whole electron free energy (zero-point energy) in the bulk or thin films is estimated to be $E_0^\mathrm{int} \sim E_0^\mathrm{sum} \sim 1$ eV.
A typical magnitude of the Casimir energy is shown in Figs.~\ref{fig:DSM}, which is $10^{2}$-$10^{4}$ times less than the whole free energy at several smaller $N_z$.
Even if the free energy itself is not observable, the Casimir effect can be measured as the thickness dependence of derivative quantities of free energy, such as internal pressure and magnetization.
For example, since magnetization is represented as the derivative of the free energy with respect to the magnetic field, its thickness dependence can signal the Casimir effect.

In general, the Casimir effect may also be induced by other quantum fields such as phonon and photon fields, but such a contribution can be distinguished from electronic one by manipulating the electron band structures using electromagnetic fields.
Also, the influence of the ``anomalous" photonic Casimir effect inside chiral medium (such as WSMs), predicted in Refs.~\cite{Fukushima:2019sjn,Canfora:2022xcx}, would be interesting, where ``anomalous" means an attractive force in the short distance and a repulsive force in the long distance (namely, the sign-flipping behavior).
Such behaviors are qualitatively different from oscillations in the fermionic Casimir effect, so that
one may distinguishably observe both the effects.
Because the typical energy scale of the Casimir effect is characterized by the speed of light for photons (modified inside semimetals) and the Fermi velocity for fermions, the fermionic Casimir effect may be smaller than the photonic one.

In addition, we comment on temperature.
The thickness dependence of thermodynamic quantities is modified by nonzero temperature (which may be called the thermal Casimir effect).
The typical energy scale of such thermal effects is proportional to its temperature.
Therefore, in experiments, if one can control low temperature of $\mu$K, the Casimir energy estimated in Figs.~\ref{fig:DSM} ($E_\mathrm{Cas} \sim$ 0.1-1 meV $\sim$ 1-10 K) will be relevant at such temperature.
At higher temperatures, the theoretical calculation of the thermal Casimir effect is straightforward in the framework of thermal field theory.

In Sec.~\ref{Sec:4}, we have investigated only the Casimir effect for the (gapless) zeroth LL, but the contributions from higher LLs can be significant in experiments.
The dispersion relations of higher LLs are described as Eq.~(\ref{eq:A6b}), where one can find that a higher LL has a gap.
Because, in general, the Casimir effect from eigenvalues with a larger gap becomes smaller compared to a gapless one, the Casimir effect from the higher LLs should be suppressed.
In this sense, the result of the zeroth LL is reliable in a strong magnetic field.
On the other hand, in a weak magnetic field, the Casimir effect from higher LLs becomes also relevant, and we have to exactly sum all the LLs for more precise predictions, which is left for future study.

%%%%%%%%%%%%%%%%%%%%%%%%%%%%%%%%
\section{Conclusion}
We have studied the Casimir effect for Dirac/Weyl fermions in Dirac/Weyl semimetals.
Our main finding is a node-induced oscillation with a period (\ref{eq:period}) in the film thickness dependence of Casimir energy.
This is induced by the existence of nodes across the Fermi level and is unknown in the Casimir effect in elementary particle physics.
Such an oscillation can be controlled by an external magnetic field and realized in thin films of realistic materials, such as Cd$_3$As$_2$ and Na$_3$Bi.
Our findings provide helpful information for developing Casimir electronics and Casimir spintronics with DSMs and WSMs.

%%%%%%%%%%%%%%%%%%%%%%%%%%%%%%%%
\section*{Acknowledgments}
The authors thank Yasufumi Araki, Tsutomu Ishikawa, and Kouki Nakata for giving us valuable comments.
This work was supported by Japan Society for the Promotion of Science (JSPS) KAKENHI (Grants No. JP17K14277 and No. JP20K14476).

  \setcounter{section}{0}
  \setcounter{equation}{0}
  \setcounter{figure}{0}
  \renewcommand{\theequation}{A\arabic{equation}}
  \renewcommand{\thesection}{A\arabic{section}}
  \renewcommand{\thefigure}{A\arabic{figure}}

\appendix
\section{Effective Hamiltonian for Weyl semimetals} \label{App:1}
In Sec.~\ref{Sec:3}, in order to demonstrate the typical behavior of the Casimir effect, we introduce typical dispersion relations for Weyl fermions in Weyl semimetals (WSMs).
Here, we write down the effective Hamiltonian~\cite{Yang:2011} on the lattice for WSMs:
\begin{align}
H^\mathrm{WSM} ( {\bf k}) =  t \sum_i^{x,y} \sigma_i \sin ak_i + \left[m - t^\prime \sum_i^{x,y,z}  (1 -\cos ak_i) \right] \sigma_z, \label{eqSP:WSM}
\end{align}
where $t$ and  $t^\prime$ are the hopping parameters, and $m$ is the onsite parameter characterizing the band structure.
This Hamiltonian breaks the time-reversal symmetry and preserves the inversion symmetry.
By diagonalizing this Hamiltonian, the two-band dispersion relations are
\begin{align}
\omega_\pm^\mathrm{WSM} ( {\bf k})  = \pm \sqrt{ t^2 \sum_i^{x,y} \sin^2 ak_i + \left[m - t^\prime \sum_i^{x,y,z} (1-\cos ak_i) \right]^2}.
\end{align}
This form is used in Sec.~\ref{Sec:3}.
Note that the momenta at the Weyl points (WPs) are given as $(k_x,k_y,k_z)=(0,0,\pm k_\mathrm{WP})$, where  $k_\mathrm{WP}= \frac{1}{a_z}\arccos{(1-\frac{m}{t^\prime})}$.

\section{Effective Hamiltonian for Weyl semimetals with a magnetic field} \label{App:2}
When the Weyl points are located near the origin ($\Gamma$ point), the Hamiltonian~(\ref{eqSP:WSM}) is reduced to be a continuous form:
\begin{align}
H^\mathrm{WSM}_\mathrm{cont} ( {\bf k}) = & t(a_xk_x \sigma_x +a_yk_y \sigma_y) \nonumber\\
&+\left[ m- \frac{t^\prime}{2} (a_x^2k_x^2+a_y^2k_y^2+a_z^2k_z^2) \right] \sigma_z,
\end{align}
where $ta_x$ and $ta_y$ can be interpreted as the Fermi velocity in the directions of $x$ and $y$, respectively.
From now on, we consider a spinless Weyl fermion, and there is no Zeeman interaction.
We replace the momenta $a_xk_x$ and $a_yk_y$ with the ladder operators $l$ and $l^\dagger$: $a_xk_x \to \sqrt{\pi \phi} (l^\dagger + l) $ and $a_yk_y \to -i \sqrt{\pi \phi} (l^\dagger - l)$.
\scriptsize
\begin{align}
&H^\mathrm{WSM}_\mathrm{cont} (k_z) 
= \nonumber\\
& \left(
\begin{array}{cc}
m- \frac{t^\prime}{2} a_z^2k_z^2 - t^\prime [2\pi \phi (l^\dagger l + \frac{1}{2})] & 2\sqrt{\pi \phi}t l \\
2\sqrt{\pi \phi}t l^\dagger & - m + \frac{t^\prime}{2} a_z^2k_z^2 + t^\prime [2\pi \phi (l^\dagger l + \frac{1}{2})]
\end{array}f
\right),
\end{align} \normalsize
where we used the commutation relation $[l,l^\dagger] =1$.
For a basis of $(0, \ket{0})^T$, the zeroth ($\nu=0$) Landau level (0LL) is obtained from the second diagonal component.
For a basis of $(\ket{\nu-1}, \ket{\nu})^T$ labeled by the index $\nu = 1,2,\cdots$, the higher Landau levels (hLLs) are obtained by diagonalizing the $2 \times 2$ Hamiltonian.
The dispersion relations for the 0LL and hLLs are given by~\cite{Nguyen:2021}
\begin{subequations}
\begin{align}
\omega^\mathrm{WSM-0LL}_\mathrm{cont}(k_z) &= -m + \frac{t^\prime}{2} a_z^2k_z^2 +\pi t^\prime \phi, \\
\omega^\mathrm{WSM-hLL}_\mathrm{cont}(k_z,\nu)  &= \pm \sqrt{K^2_{\nu k_z} + 4\pi t^2 \phi \nu} +\pi t^\prime \phi, \label{eq:A5b} \\
K_{\nu k_z} &= m-\frac{t^\prime}{2} a_z^2k_z^2 -2\pi t^\prime \phi \nu.
\end{align}
\end{subequations}
Again we replace $k_z$ as $a^2k_z^2 \to 2(1-\cos ak_z)$, the final forms are~\cite{Nguyen:2021}
\begin{subequations}
\begin{align}
\omega^\mathrm{WSM-0LL}(k_z) &= -m + t^\prime (1-\cos ak_z) + \pi t^\prime \phi, \label{eq:WSM-0LL_lat} \\
\omega^\mathrm{WSM-hLL}(k_z,\nu)  &= \pm \sqrt{K^2_{\nu k_z} + 4\pi t^2 \phi \nu} +\pi t^\prime \phi, \label{eq:A6b} \\
K_{\nu k_z} &= m-t^\prime(1-\cos ak_z) -2\pi t^\prime \phi \nu.
\end{align}
\end{subequations}
This form for the 0LL is used in Sec.~\ref{Sec:4}.

%%%%%%%%%%%%%%%%%%%%%%%%%%%%%%%%
%\bibliographystyle{cas-model2-names}
\bibliographystyle{apsrev4-1}
\bibliography{ref}

  \setcounter{section}{0}
  \setcounter{equation}{0}
  \setcounter{figure}{0}
  \renewcommand{\theequation}{S\arabic{equation}}
  \renewcommand{\thesection}{S\arabic{section}}
  \renewcommand{\thefigure}{S\arabic{figure}}

\onecolumn
\section*{Supplementary material for ``Dirac/Weyl-node-induced oscillating Casimir effect"}
%\tableofcontents
\section{Effective Hamiltonian for Dirac semimetals and additional analyses with strained Dirac semimetals}
In the main text, we investigate the Casimir effects for Dirac fermions by focusing on systems with Dirac points near the Fermi level.
In this supplementary material, we investigate whether the Casimir effect is affected by (i) the position of the Fermi level or (ii) gaps opening at the Dirac points.
In order to study this, we use the model parameters in Ref.~\cite{Arribi:2020}, where the model parameters for unstrained Cd$_3$As$_2$ are estimated by fitting the numerical results from the density-functional theory.

A low-energy effective Hamiltonian for Dirac fermions in Dirac semimetals (DSMs) was proposed by Wang {\it et al.}~\cite{Wang:2012,Wang:2013}:
In the four-band basis of $\ket{S_{\frac{1}{2}},\frac{1}{2}}$, $\ket{P_{\frac{3}{2}},\frac{3}{2}}$, $\ket{S_\frac{1}{2},-\frac{1}{2}}$, and $\ket{P_\frac{3}{2},-\frac{3}{2}}$,
\begin{subequations}
\begin{align}
H^\mathrm{DSM} ( {\bf k})
&=
\left(
\begin{array}{cccc}
\epsilon_0 ({\bf k}) + M ({\bf k}) & A( k_x+ik_y) & D( k_x-ik_y)  & B^\ast ({\bf k}) \\
A(k_x-ik_y)  &\epsilon_0 ({\bf k}) -M ({\bf k}) &  B^\ast ({\bf k}) & 0\\
D( k_x+ik_y)  &   B ({\bf k}) & \epsilon_0 ({\bf k}) +M ({\bf k})& -A(k_x-ik_y)\\
B ({\bf k}) & 0 & -A(k_x+ik_y) & \epsilon_0 ({\bf k}) -M ({\bf k}) \\
\end{array}
\right), \\
\epsilon_0  (\bf{k}) &= C_0 + C_1k_z^2 + C_2(k_x^2+k_y^2), \label{eq:DSM_epsilon} \\
M  (\bf{k}) &= M_0 +M_1k_z ^2 +M_2 (k_x^2+k_y^2).
\end{align}
\end{subequations}
Here, $D$ is a parameter characterizing the inversion symmetry breaking, and below we set $D=0$.
When we introduce $B=b_1 k_z$ which breaks the fourfold rotational $C_4$ symmetry around the $z$ axis for Cd$_3$As$_2$, the two Weyl nodes in a Dirac node are coupled with each other, and a gap is opened~\cite{Wang:2012,Wang:2013}.
In Ref.~\cite{Arribi:2020}, this term is related to a strain, where $b_1 >0$ ($b_1 <0$) induces a compressive (tensile) strain along the $x$ axis.
By diagonalizing the Hamiltonian, the four-band dispersion relations are given by
\begin{align}
\omega_\pm^\mathrm{DSM} ( {\bf k})  = \epsilon_0 \pm \sqrt{M^2+A^2(k_x^2+k_y^2) + B^2}, \label{eq:DSMb}
\end{align}
where the spin-up and spin-down bands are degenerate.
We replace the momentum as $k_i^2 \to \frac{1}{a_i^2} \sin^2 ak_i$ for the term proportional to $A$ or $B$ and $k_i^2 \to \frac{1}{a_i^2} (2-2\cos ak_i)$ for the other terms.
The model parameters estimated in Ref.~\cite{Arribi:2020} are summarized in Table.~\ref{tab:model_para}, where $b_1$ corresponds to the compressive strain of $-0.7$\% along the $x$ axis.

\begin{table}[bh!]
  \centering
  \caption{Model parameters for strained Cd$_3$As$_2$ estimated in Ref.~\cite{Arribi:2020}.
We also show the other parameters for unstrained Cd$_3$As$_2$~\cite{Cano:2016eie} and Na$_3$Bi~\cite{Wang:2012} which are used in our analysis in the main text.}
\begin{tabular}{l c c c c} 
  \hline \hline
Parameters   &  Strained Cd$_3$As$_2$~\cite{Arribi:2020} & Unstrained Cd$_3$As$_2$~\cite{Cano:2016eie} & Na$_3$Bi~\cite{Wang:2012} \\ 
  \hline
$A$ (eV\AA)            & $1.089$   & $0.889$     & $2.4598 $ \\
$C_0$ (eV)               & $0.0113$ & $ -0.0145$ & $-0.06382$ \\
$C_1$ (eV\AA$^2$)  & $12.05$   & $10.59$     & $8.7536$  \\
$C_2$ (eV\AA$^2$)  & $13.13$   & $11.5$       & $ -8.4008$ \\
$M_0$ (eV)              & $0.0374$  & $ -0.0205$ & $ -0.08686$\\
$M_1$ (eV\AA$^2$)  &  $-20.36$  & $18.77$     & $10.6424$ \\
$M_2$ (eV\AA$^2$)  &  $-18.77$  & $13.5$      & $10.361$ \\
$b_1$ (eV\AA)         & $0.2566$  & $0$          & $0$ \\
$a_x=a_y$ (\AA)        &  $12.633$  & $12.67$     & $ 5.448$ \\
$a_z$ (\AA)              &  $25.427$  & $25.48$     & $9.655$ \\
  \hline \hline   
  \end{tabular}
  \label{tab:model_para}
\end{table}

\begin{figure}[t!]
    \centering
    \begin{minipage}[t]{0.25\columnwidth}
    \includegraphics[clip,width=1.0\columnwidth]{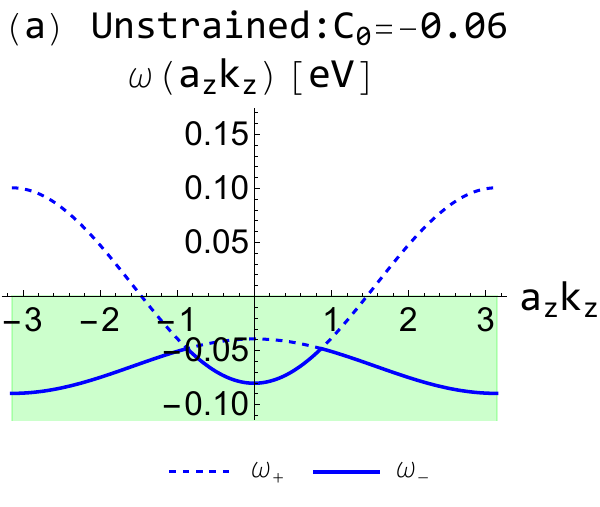}
    \end{minipage}%
    \begin{minipage}[t]{0.25\columnwidth}
    \includegraphics[clip,width=1.0\columnwidth]{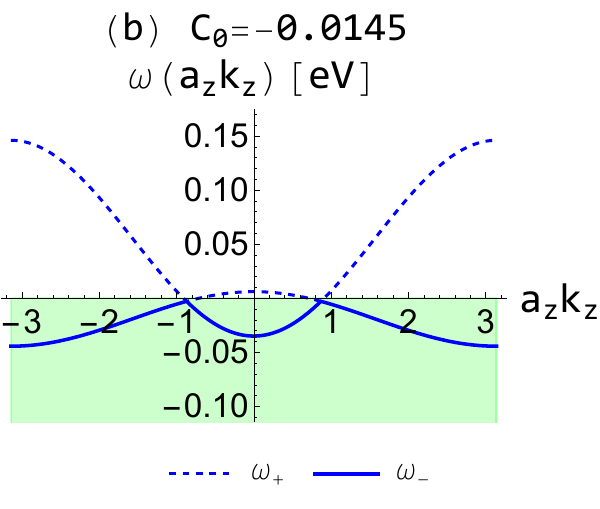}
    \end{minipage}%
    \begin{minipage}[t]{0.25\columnwidth}
    \includegraphics[clip,width=1.0\columnwidth]{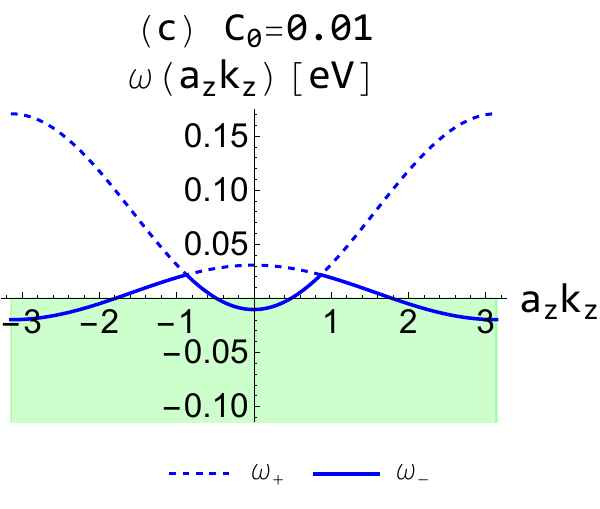}
    \end{minipage}
    \begin{minipage}[t]{0.25\columnwidth}
    \includegraphics[clip,width=1.0\columnwidth]{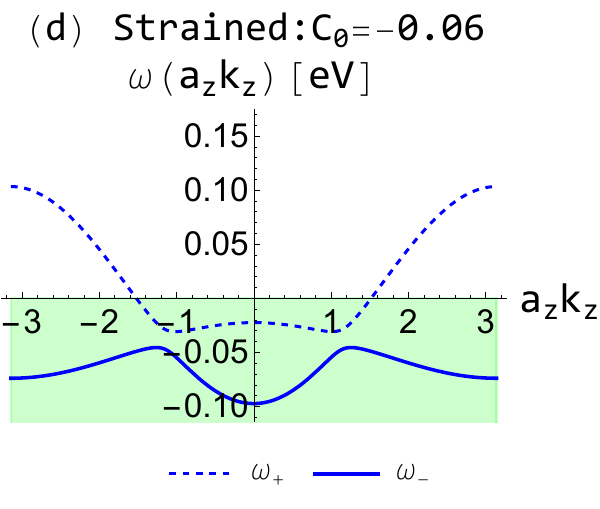}
    \end{minipage}%
    \begin{minipage}[t]{0.25\columnwidth}
    \includegraphics[clip,width=1.0\columnwidth]{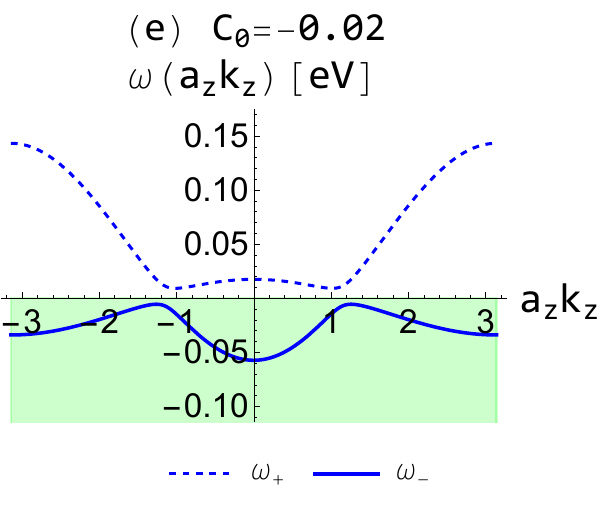}
    \end{minipage}%
    \begin{minipage}[t]{0.25\columnwidth}
    \includegraphics[clip,width=1.0\columnwidth]{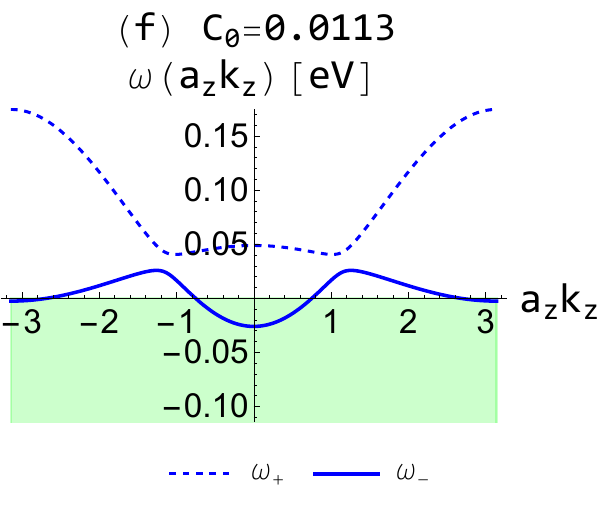}
    \end{minipage}
    \begin{minipage}[t]{0.5\columnwidth}
    \includegraphics[clip,width=1.0\columnwidth]{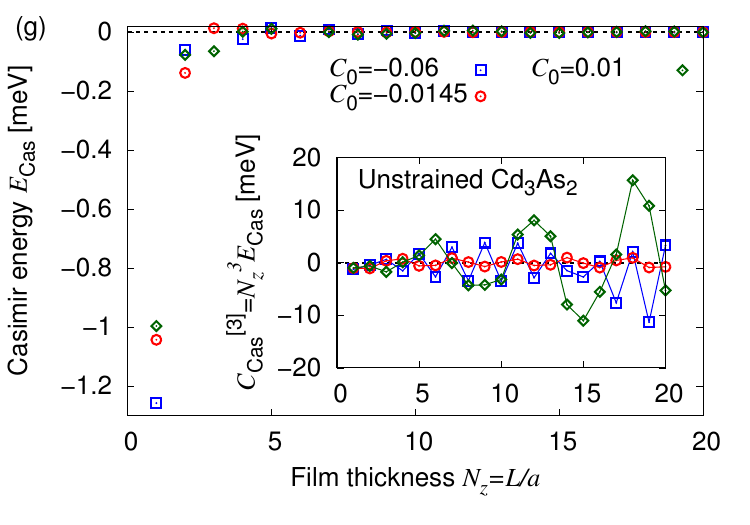}
    \end{minipage}%
    \begin{minipage}[t]{0.5\columnwidth}
    \includegraphics[clip,width=1.0\columnwidth]{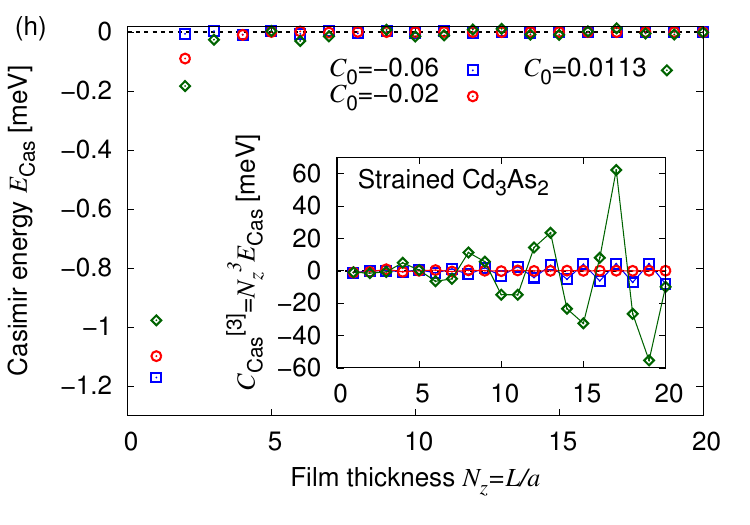}
    \end{minipage}
    \caption{Dispersion relations, Casimir energy, and Casimir coefficient for Dirac electrons in Cd$_3$As$_2$ thin films, described by the model (\ref{eq:DSMb}).
(a)-(f) Dispersion relations in the $z$ direction.
(g) Casimir energy and Casimir coefficient for {\it unstrained} Cd$_3$As$_2$ thin films.
(h) Casimir energy and Casimir coefficient for {\it strained} Cd$_3$As$_2$ thin films.
 }
    \label{strain}
\end{figure}

In Figs.~\ref{strain}, we show the Fermi level shifted dispersion relations [(a)-(f)] and the corresponding Casimir energies [(g) and (h)].
The position of the Fermi level can be tuned by the parameter $C_0$, and we show some examples with different $C_0$.
The unstrained case at $C_0=-0.0145$ [Fig.~\ref{strain}(b)] estimated in Ref.~\cite{Cano:2016eie} is the same as the result shown in the main text, where the period of oscillation is $\tau_\mathrm{Cas} \sim 3.5$.
If we change $C_0$ to $-0.06$ [Fig.~\ref{strain}(a)] or $0.01$ [Fig.~\ref{strain}(c)], then the position of the Fermi level is shifted, and the Casimir energy is modified. 
At $C_0=-0.06$, the position of the Fermi level becomes higher.
As a result, $\tau_\mathrm{Cas}$ becomes longer.
This change can be understood by the formula $\tau_\mathrm{Cas} \sim \pi/a_z k_\mathrm{FP}$ proposed in the main text:
The two Fermi points are located at $a_z k_\mathrm{FP} \sim \pm \pi/2 $, so that $\tau_\mathrm{Cas} \sim 2$, which is consistent with the numerical result.
At $C_0=0.01$, the position of the Fermi level becomes lower.
As a result, $\tau_\mathrm{Cas} \sim 6$-$7$.
Thus, the shift of the Fermi level can tune the period of oscillation through the shifts of the Fermi points.
Experimentally, one can tune the positions of the Fermi points by doping electrons/holes or applying gate voltages.

Next, we discuss the effect of gaps in Dirac nodes.
For the strained case at $C_0=-0.02$ [Fig.~\ref{strain}(e)], the gaped nodes are located near the Fermi level.
Compared to the unstrained case at $C_0=-0.0145$, the magnitude of $E_\mathrm{Cas}$ becomes smaller due to the gap effect.
The suppression of $E_\mathrm{Cas}$ is similar to the typical behavior of the Casimir effect for massive fields.
Thus, a gap tends to weaken the Casimir effect, and the period of oscillation is unchanged.
At $C_0=-0.06$, we find an oscillation with $\tau_\mathrm{Cas} \sim 2$, which is similar to the unstrained case.
At $C_0=0.0113$ estimated in Ref.~\cite{Arribi:2020}, $E_\mathrm{Cas}$ becomes larger than the unstrained case at $C_0=0.01$.
Thus, one can control the Casimir effect by straining materials. 

We comment on the cases of fully occupied bands ($C_0$ is small enough) or fully unoccupied bands ($C_0$ is large enough).
In these situations, we find that the Casimir energy is exactly zero.
This is because the integrand of the zero-point energy is $|\omega_+^\mathrm{DSM}| + |\omega_-^\mathrm{DSM}| = 2\epsilon_0$ using the dispersion relations~(\ref{eq:DSMb}), and only the terms~ (\ref{eq:DSM_epsilon}) with a constant and a quadratic dispersion survives.
This is the same mechanism as the Casimir effect for quadratic-like dispersions on the lattice, proposed in Ref.~\cite{Nakayama:2022ild}.

\section{Additional analyses with periodic boundary conditions}
In the main text, we have considered thin films with a phenomenological boundary condition, where the $z$-component of the momentum is discretized as $a_zk_z \to \frac{n\pi}{N_z}$ ($n=1,\cdots, 2N_z$) (See the main text). 
As another instructive example to demonstrate the Casimir effect, hereafter we study the periodic boundary condition, where $ak_z \to \frac{2\pi n}{N_z}$ ($n=0,1,\cdots,N_z-1$).

\begin{figure}[h!]
    \centering
    \begin{minipage}[t]{0.5\columnwidth}
    \includegraphics[clip,width=1.0\columnwidth]{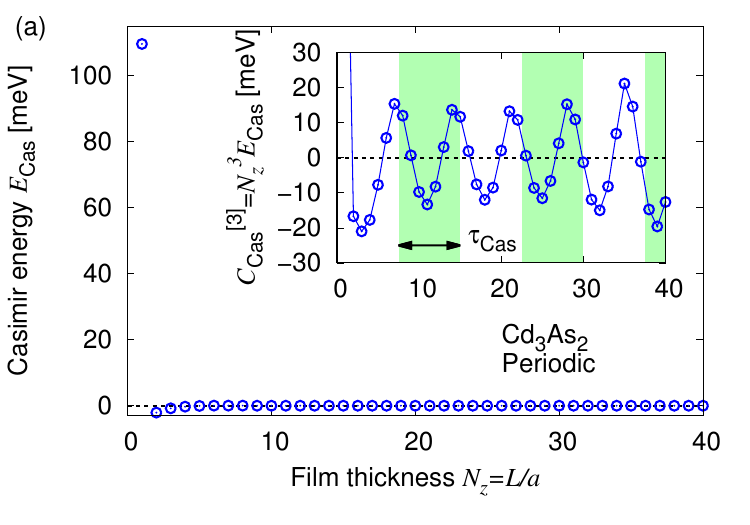}
    \end{minipage}%
    \begin{minipage}[t]{0.5\columnwidth}
    \includegraphics[clip,width=1.0\columnwidth]{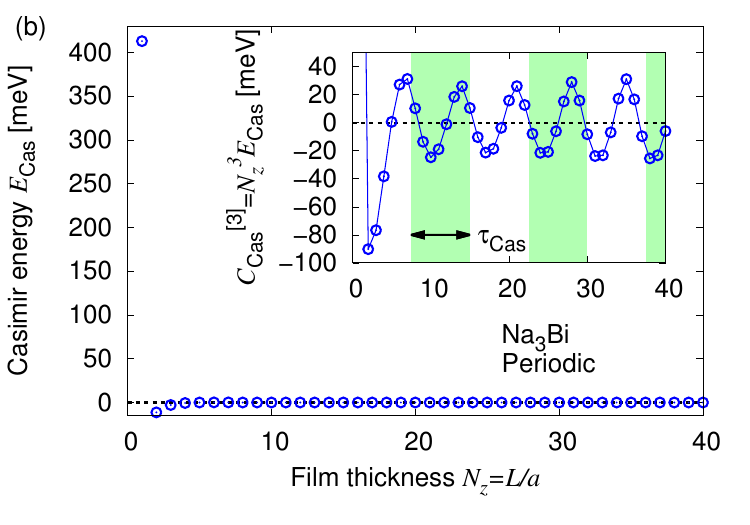}
    \end{minipage}
    \caption{Casimir energy and Casimir coefficient for Dirac electrons in (001)-oriented thin films of Dirac semimetals described by the model (\ref{eq:DSMb}) with {\it the periodic boundary condition}.
(a): Cd$_3$As$_2$.
(b): Na$_3$Bi. 
Inset: the period of $\tau_\mathrm{Cas} \sim 7$ is shown as the colored or uncolored regions.}
    \label{fig:DSM_PBC}
\end{figure}

The numerical results are shown in Figs.~\ref{fig:DSM_PBC}(a) and (b) for Cd$_3$As$_2$ and Na$_3$Bi, respectively, where the model parameters for the dispersion relations are the same as the values in the main text (and also in Table.~\ref{tab:model_para}).
From these figures, we find the oscillating Casimir effect, and its period is given by the following formula:
\begin{align}
\tau_\mathrm{Cas}^\mathrm{PBC} = \frac{2\pi}{a_z k_\mathrm{WP/DP}}.
\end{align}
The only difference between the formulas for the periodic boundary and the phenomenological boundary is the factor of $2$.
This is because the number of eigenvalues at a fixed $N_z$ changes as $2N_z \to N_z$.
Hence, the period for the periodic boundary is longer than that for the phenomenological boundary.
As a result, the periods for Cd$_3$As$_2$ and Na$_3$Bi with the periodic boundary are estimated as $\tau_\mathrm{Cas} \sim 7$.
Furthermore, we also find that the magnitude of the Casimir energy with the periodic boundary is larger than that with the phenomenological boundary.
This is because $2N_z \to N_z$ increases the difference between the integral part and the sum part of the zero point energy.

\section{Additional analyses with zeroth Landau levels}
\begin{figure}[t!]
    \centering
    \begin{minipage}[t]{0.5\columnwidth}
    \includegraphics[clip,width=1.0\columnwidth]{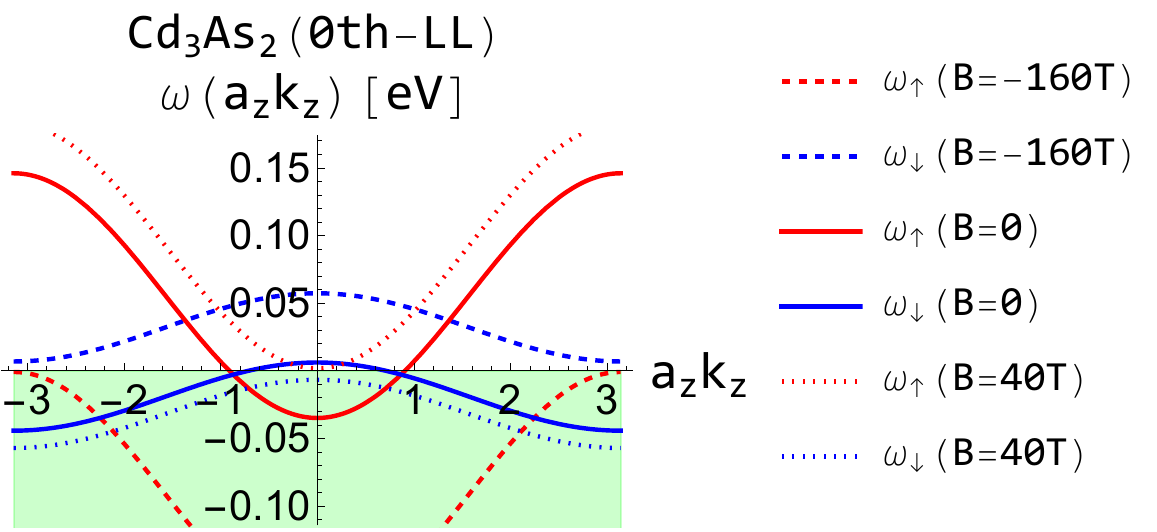}
    \end{minipage}
    \caption{
Dispersion relations of the zeroth Landau levels (0LLs) of Cd$_3$As$_2$ thin films, defined as Eqs.~(\ref{eq:spinup2}) and (\ref{eq:spindown2}).
}
    \label{fig:disp_0LL}
\end{figure}

In the main text, we have shown qualitative properties of Casimir energy in Weyl/Dirac semimetals in magnetic fields.
Here, we demonstrate a quantitative analysis.
In this section, for simplicity, we use the following dispersion relations for the zeroth Landau levels of a Dirac semimetal with the Zeeman coupling (analyses for higher Landau levels are also straightforward):
For a magnetic field $B$ parallel to the $z$ direction~\cite{Nguyen:2021},
\begin{subequations}
\begin{align}
\omega^\mathrm{DSM-0LL}_\uparrow
&=
C_0 + M_0
+\frac{C_1 + M_1}{a_z ^2}(2 - 2\mathrm{cos} a_zk_z)
+\left[(C_2 + M_2)\frac{2\pi e}{h} + \frac{\mu_B g_\uparrow}{2} \right]B, \label{eq:spinup2} \\
\omega^\mathrm{DSM-0LL}_\downarrow
&=
C_0 - M_0
+\frac{C_1 - M_1}{a_z ^2}(2 - 2\mathrm{cos} a_zk_z)
+\left[(C_2 - M_2)\frac{2\pi e}{h} - \frac{\mu_B g_\downarrow}{2}\right] B. \label{eq:spindown2}
\end{align}
\end{subequations}
For model parameters, we use those given as ``unstrained Cd$_3$As$_2$" in Table.~\ref{tab:model_para}.
The other physical constants are known as the Planck constant $h = 6.62607015\times 10^{-34}$ J$\cdot$s, the elementary charge $e = 1.602176634\times 10^{-19}$ C, and the Bohr magneton $\mu_B = 9.2740100783 \times 10^{-24}$ JT$^{-1}$.
We also assume effective $g$ factors as $g_\uparrow = 18.6$ and $g_\downarrow = 10.0$ ($g_\uparrow$ is an estimate in Ref.~\cite{Jeon:2014} and $g_\downarrow$ is a simplified assumption in Ref.~\cite{Nguyen:2021}).
In Fig.~\ref{fig:disp_0LL}, we show three examples of dispersion relations at fixed magnetic fields: $B=0$, $-160$ T, and $40$ T.

\begin{figure}[tb!]
    \centering

    \begin{minipage}[t]{0.5\columnwidth}
    \includegraphics[clip,width=1.0\columnwidth]{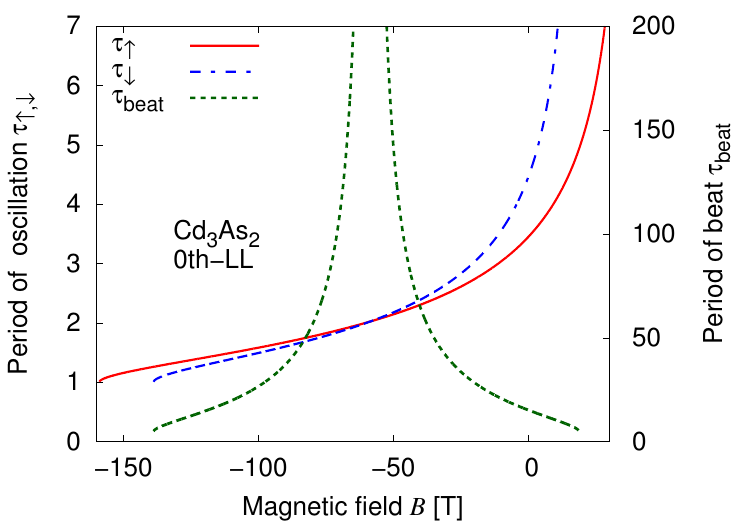}
    \end{minipage}
    \caption{
Magnetic field dependences of oscillation periods in the Casimir energy originated from the zeroth Landau levels (0LLs) of Cd$_3$As$_2$ thin films.
A difference between two periods, $\tau_\uparrow$ and $\tau_\downarrow$ for the up-spin and down-spin bands, leads to a beat with a period $\tau_\mathrm{beat}$.
}
    \label{fig:0LL}
\end{figure}

Using Eqs.~(\ref{eq:spinup2}) and (\ref{eq:spindown2}), we quantitatively discuss the properties of the Casimir energy in a wide range of magnetic fields.
Here, we adopt the phenomenological boundary condition, $a_zk_z \rightarrow \frac{n\pi}{N_z}\ (n = 1,...,2N_z)$ and $\sum_n \rightarrow \frac{1}{2}\sum_{n}$, in the definition of the Casimir energy.

In Fig.~\ref{fig:0LL}, we show the magnetic field dependence of the oscillation period $\tau_\mathrm{Cas} = \pi/a_zk_\mathrm{FP}$, where we have newly defined $\tau_\uparrow$ from the up-spin band and $\tau_\downarrow$ from the down-spin band.
First we note the region of very strong magnetic fields, such as $B< -158.9$ T and $B> 38.2$ T for the up spin and $B< -138.9$ T and $B> 18.8$ T for the down spin.
In this region, the Casimir energies are exactly zero (that is, there is no oscillation) within our model.
This is because each band does not cross the Fermi level, and $|\omega^\mathrm{DSM-0LL}_{\uparrow,\downarrow}|$ becomes a constant plus cosine function, which is formally the same as the mechanism proposed in Ref.~\cite{Nakayama:2022ild}.
In magnetic fields smaller than such a ``threshold," we can find the nonzero oscillating Casimir energy and the magnetic field dependence of $\tau_\uparrow$ and $\tau_\downarrow$.
By definition, these periods are distributed within in the range of $1< \tau_\mathrm{Cas}< \infty$, which corresponds to the positions of the Fermi points $\pi > a_zk_\mathrm{FP} > 0$.
We also find that the $B$-dependence of $\tau_\downarrow$ is stronger than $\tau_\uparrow$, which is determined by the $B$-dependence of its band structure.
Furthermore, from Fig.~\ref{fig:0LL}, we find the two lines of $\tau_\uparrow$ and $\tau_\downarrow$ intersect ($\tau_\downarrow = \tau_\uparrow$) at $B=-58.6$ T, which means that the period of the beat, defined as $\tau_\mathrm{beat} = \tau_\uparrow\tau_\downarrow/|\tau_\uparrow - \tau_\downarrow|$, diverges at the same time.
We can also find the period of the beat is limited to $\tau_\mathrm{beat} > 5$.
Such a lower bound of $\tau_\mathrm{beat}$ is determined by the details of the band structure.

Finally, we comment on the small-magnetic-field region, particularly, Eqs.~(\ref{eq:spinup2}) and (\ref{eq:spindown2}) at $B=0$.
Eqs.~(\ref{eq:spinup2}) and (\ref{eq:spindown2}) are the dispersion relations of the zeroth Landau levels, but the band structure composed of two eigenvalues at $B=0$ has the same form as that of the $z$ direction of the original dispersion relations of Cd$_3$As$_2$ [defined as Eq.~(\ref{eq:DSMb}) at $b_1=0$] without magnetic fields.
The differences are the factor of spin degeneracy/splitting and the dispersion in the $x$ and $y$ directions.
It should be noted that these differences do not change the qualitative property of the Casimir effect (induced by the momemtum discretization in the $z$ direction), which indicates that the Casimir effect from the Eqs.~(\ref{eq:spinup2}) and (\ref{eq:spindown2}) at $B=0$ is analogous to that from Eq.~(\ref{eq:DSMb}). 
As an example, using $\tau_\mathrm{Cas} = \pi/a_zk_\mathrm{FP}$, from Eqs.~(\ref{eq:spinup2}) and (\ref{eq:spindown2}) at $B=0$, we obtain the oscillation periods $\tau_\uparrow \sim 3.45$ and $\tau_\downarrow \sim 4.46$, respectively.
These values are consistent with the rough estimate using the position of the Dirac points of Cd$_3$As$_2$, $\tau_\mathrm{Cas} =
\pi/a_zk_\mathrm{DP} \sim \pi/\sqrt{-a_z ^2 M_0/M_1} \sim 3.73$.

\section{Analytic solutions of Casimir energy}
In the main text, we have shown the numerical results of the Casimir energy, but in some cases, we can obtain their analytic solutions. 
Here, we give analytic solutions of the Casimir energy for a simplified dispersion relation.

We define the Casimir energy for a one-dimensional spinless fermion system with the phenomenological boundary condition, $a_zk_z \rightarrow \frac{n\pi}{N_z}\ (n = 1,...,2N_z)$:
\begin{subequations}
\begin{align}
E_\mathrm{Cas}(N_z)
&\equiv  E_0^\mathrm{sum}(N_z) -E_0^\mathrm{int}(N_z), \label{eq_sup:Ecas}\\
E_0^\mathrm{sum}(N_z)
&=
-\frac{1}{2}
\times
\frac{1}{2}
\sum_{n=1} ^{2N_z}
|\omega_n|, \label{eq_sup:E0sum}\\
E_0^\mathrm{int}(N_z)
&=
-\frac{N_z}{2}
\int_{0} ^{2\pi} \frac{d(a_zk_z)}{2\pi}
|\omega (k_z)|. \label{eq_sup:E0int}
\end{align}
\end{subequations}
Here, as an instructive example, we consider the following dispersion relations:\footnote{Note that this dispersion relation is regarded as a one-dimensional cosine band and also is formally the same as the lowest Landau level of three-dimensional time-reversal symmetry breaking Weyl semimetals.
As other instructive examples, when the two dispersion relations of a one-dimensional Weyl fermions (or equivalently the lowest Landau levels of three-dimensional Dirac fermions) are given as $ \omega_\pm^\mathrm{1dWSM} = \pm \omega^\mathrm{1d-cos}$, and its Casimir energy becomes two times larger than that for $\omega^\mathrm{1d-cos}$.
}
\begin{align}
\omega^\mathrm{1d-cos} (k_z)
=
C -  \mathrm{cos}(a_zk_z),
\end{align}
where a constant parameter $C$ characterizes the positions of Fermi points (FPs).
In particular, in the range of $-1<C<1$, this band crosses the Fermi level at $a_zk_z = \pm a_z k_\mathrm{FP}= \pm \arccos C$: For example, when we fix the Fermi points at $a_zk_\mathrm{FP} = \pi/2, \pi/3, \pi/4$, and $\pi/6$, the parameter $C$ are tuned as $0,1/2, 1/\sqrt{2}$, and $\sqrt{3}/2$, respectively.

Since $N_z \geq 1$ is an integer, the analytic solutions of the Casimir energy are
\begin{align}
&E_\mathrm{Cas}(N_z,C) = \nonumber\\
&\left\{ \begin{aligned}
&0 \ \ \ (\mathrm{for} \ C \leq -1 \ \mathrm{or} \ C \geq 1), \\
&\frac{\left( \sqrt{1-C^2} + C \arcsin C \right) N_z}{\pi}  -\frac{1}{2} \times \\
&\left[ C \left(1+N_z -2 \mathrm{Ceil}\left[ \frac{N_z\arccos C}{\pi} \right] \right) 
- \mathrm{Csc} \left(\frac{\pi}{2N_z}\right)
\sin \left(\frac{\pi - 2\pi \mathrm{Ceil} \left[\frac{N_z \arccos C}{\pi} \right]}{2N_z}\right)  \right] \ \ \ (\mathrm{for} \ -1 \leq C \leq 1). 
\end{aligned} \right.
\label{eq_sup:anal_nonzero}
\end{align}
Thus, for $C \leq -1$ or $C \geq 1$, the Casimir energy is exactly zero because the cosine band does not cross the Fermi level (see Ref.~\cite{Nakayama:2022ild}).
On the other hand, the band structures with $-1<C<1$ induce the nonzero Casimir energy.
The first term comes from the integral part (\ref{eq_sup:E0int}), which is proportional to $N_z$ by definition.
The second term corresponds to the sum part (\ref{eq_sup:E0sum}).
The existence of the ceiling functions $\mathrm{Ceil}(a_zk_\mathrm{FP}N_z/\pi)$ indicates the oscillation of the Casimir energy with a period $\tau_\mathrm{Cas} = \pi/a_zk_\mathrm{FP}$.

%\bibliographystyle{utphys}
%\bibliographystyle{apsrev4-1}
%\bibliography{ref}

\end{document}